\documentclass[doublecol]{epl2} 
\title{Dynamic glass transition: bridging the gap between mode-coupling theory
and the replica approach}
\shorttitle{Dynamic glass transition: MCT and replica approach}

\author{Grzegorz Szamel}
\shortauthor{G. Szamel}

\institute{Department of Chemistry, 
Colorado State University, Fort Collins, CO 80523}


\abstract{We clarify the relation between the ergodicity breaking transition predicted
by mode-coupling theory and the so-called dynamic transition 
predicted by the static replica approach. Following 
Franz and Parisi [Phys. Rev. Lett. \textbf{79}, 2486 (1997)], 
we consider a system of particles in a metastable state characterized 
by non-trivial correlations with a quenched configuration. We show that the 
assumption that in a metastable state particle currents vanish leads to 
an expression for the replica off-diagonal direct correlation 
function in terms of a non-trivial part of the 
replica off-diagonal static four-point correlation function. 
A factorization approximation for this function results in an approximate closure 
for the replica off-diagonal direct correlation function. The replica off-diagonal
Ornstein-Zernicke equation combined with this closure coincides with 
the equation for the non-ergodicity parameter derived using the mode-coupling theory.}

\begin{document}

\maketitle

\section{Introduction}

When trying to understand a new physical phenomenon one usually starts 
with a mean field theory and only afterwards 
one tries to account for fluctuation effects. 
The glass transition has been studied
for many decades but theories which aspire to describe it at a mean field level,
starting from a microscopic description of an interacting many-particle system,
are quite recent. First, there is mode-coupling theory,
derived in early eighties by G\"otze and collaborators \cite{BGS,Goetze}. 
Second, there is the replica approach formulated in the late nighties 
by Mezard, Franz and Parisi \cite{MP1,FP,CFP,MP2}. While the 
latter theory makes references to the former, the precise relation between them
is unclear (with the exception of disordered spin models). 
Our goal in this Letter is to bridge the gap between these theories.

The mode-coupling theory of the glass transition is, essentially, a self-consistent 
generalization of earlier mode-coupling approaches dealing with long-wavelength
dynamics near critical points \cite{critical} and long-time tails
in time-dependent correlation functions \cite{EHvL} to the dynamics
of supercooled fluids at all length and time scales. The fundamental
approximation of this theory consists in a factorization of a 
time-dependent four-point correlation function evolving with the 
so-called projected dynamics \cite{Goetze} (or, in the case of an overdamped
system, with the so-called irreducible dynamics \cite{SL}). 
The somewhat mysterious nature of the four-point
function involved, the uncontrolled factorization approximation,
and the difficulty to calculate corrections have made the mode-coupling theory quite   
controversial. On the other hand, this 
theory makes a number of predictions \cite{Goetze} 
using as the only input the static structure of the system. Thus, it
is verifiable and, therefore, it inspired a number of experimental and
simulational studies. The major conclusion reached in pioneering simulations 
of Hansen and collaborators \cite{Hansen} 
is that the ergodicity breaking transition predicted by the
mode-coupling theory corresponds to a dynamic crossover (originally 
named the kinetic glass transition) occurring at a density smaller than
(or at a temperature larger than) that of the laboratory glass transition. 
Later simulations tested, \textit{inter alia}, the wave-vector
dependence of the so-called non-ergodicity parameter \cite{nep} and found a remarkable
agreement with the predictions of the mode-coupling theory \cite{SK}.

In contrast to the mode-coupling theory, the replica approach, as 
formulated by Mezard, Franz and Parisi \cite{MP1,FP,CFP,MP2} 
and further developed by Parisi and Zamponi \cite{PZ1,PZ2}, 
is a purely static theory. 
It is rooted in the analysis of disordered spin models with long-range
interactions (solvable exactly within mean-field theory) which exhibit
the so-called one-step replica symmetry breaking. As showed in the remarkable
series of papers by Kirkpatrick, Thirumalai and Wolynes \cite{KTW}, these models 
have two transitions, a dynamic transition at which time-dependent correlation
functions stop decaying and a static transition at which the co-called configurational
entropy (the logarithm of the number of metastable glassy states) vanishes. 
The dynamics of these models is described exactly by equations
identical to the so-called schematic mode-coupling equations. The dynamic
transition coincides with the ergodicity breaking transition predicted by 
the latter equations. The replica theory of Mezard, Parisi and collaborators
builds upon Kirkpatrick \textit{et al.}'s contribution and 
deals with the thermodynamic properties of supercooled many-particle systems
without quenched disorder. It also predicts the existence
of two transitions: the so-called dynamic transition at which the phase
space of the system becomes split into a collection of metastable glassy states
and the so-called ideal glass transition at which the configurational entropy
vanishes. The former transition is identified with the ergodicity-breaking
transition predicted by the mode-coupling theory. 

It should be recalled at this point that the mode-coupling theory 
was not proposed as a mean-field theory of the glass transition.
In fact, this theory descends from
approaches that were derived to describe distinctly non-mean-field features of 
many-particle dynamics \cite{critical,EHvL}. 

In addition, there is another somewhat unclear aspect of the identification of 
the ergodicity breaking transition of the mode-coupling theory and the dynamic
transition of the replica approach: both the locations of these transitions 
and the frozen-in correlations associated with them are markedly different. 
In particular, while the non-ergodicity parameter 
predicted by the mode-coupling
theory agrees well with simulations \cite{SK}, the one inferred \cite{PZ2} 
from the replica approach
does not \cite{MP1,PZ2}. Recently, it has been showed that the difference between the 
locations of the two transitions increases 
strongly with increasing spatial dimension \cite{SS,IM}. 
This fact, together with a surprising behavior
of non-ergodicity parameters in high spatial dimensions, prompted Ikeda and Miyazaki
\cite{IM} to question the validity of the mode-coupling theory. 

The goal of this Letter is to clarify the relationship between the mode-coupling
theory and the replica approach. We start with recalling that the
replica approach is a general scheme that, in order to make quantitative 
predictions, has to be accompanied by specific approximations of the liquid
state theory. Early versions of the replica approach
\cite{MP1,CFP} used the so-called replicated hyper-netted chain approximation.
Later versions \cite{MP2,PZ1} used a molecular version of this
approximation combined with the so-called small cage radius expansion. Finally,
the most recent version \cite{PZ2} used an effective potential approach 
combined with the same small cage radius expansion. It was argued
\cite{MP2,PZ2} that the latter two approaches are well suited to describe
dense glassy systems and the ideal glass transition. However, they are
not appropriate to describe the dynamic transition, except in the limit of large
spatial dimension \cite{PZ2}. 
Here we propose an alternative to the hyper-netted chain approximation used in 
Refs. \cite{MP1,CFP} that predicts the dynamic transition that coincides with 
the mode-coupling transition \cite{KW}.

\section{Replica approach} 
We follow Franz and Parisi \cite{FP,CFP}: We consider a system
of particles which is tightly pinned to a quenched configuration. We imagine
that the pinning potential is gradually relaxed and that at each step of 
this process the 
system is allowed to equilibrate (according to the Boltzmann measure 
including the pinning potential). The main assumption is that, if the 
density is high enough (or if the temperature is low enough) at a certain point 
the system gets stuck in a metastable state characterized by non-trivial 
correlations with the quenched configuration. We assume that this
metastable state survives in the limit of vanishing pinning potential. 
Finally, we assume self-averaging and we
average over a distribution of quenched configurations using 
the equilibrium Boltzmann measure. To perform this average 
we replicate the system and at the end of the analysis we take 
the limit of the number of replicas of the system $r$ going to zero.
Including the quenched configuration, we have $m=r+1$ replicas and we take
the limit $m\to 1$. 

\section{Ornstein-Zernicke (OZ) equations}  
In a metastable state 
we can use the standard apparatus of equilibrium statistical mechanics,
including replicated OZ equations,
\begin{eqnarray}
\label{rOZ}
\delta n_{\alpha\beta}(\mathbf{r}_1,\mathbf{r}_2) &=&
n^2 c_{\alpha\beta}(\mathbf{r}_1,\mathbf{r}_2) 
\\ \nonumber && + 
n \sum_{\gamma} \int d\mathbf{r}_3 c_{\alpha\gamma}(\mathbf{r}_1,\mathbf{r}_3)
\delta n_{\gamma\beta}(\mathbf{r}_3,\mathbf{r}_2). 
\end{eqnarray}
Here $\alpha$, $\beta$ and $\gamma$ are replica indices, 
$\delta n_{\alpha\beta}(\mathbf{r}_1,\mathbf{r}_2)$ is the nontrivial part of
the two-particle density, 
$\delta n_{\alpha\beta}(\mathbf{r}_1,\mathbf{r}_2)=
n_{\alpha\beta}(\mathbf{r}_1,\mathbf{r}_2)-n^2$, 
$n$ is the single-particle density (assumed to be uniform and 
the same in all replicas) and $c_{\alpha\beta}(\mathbf{r}_1,\mathbf{r}_2)$
is the direct correlation function. Note that Eqs. (\ref{rOZ}) essentially define 
$c_{\alpha\beta}$'s 
and additional assumptions are needed
in order to make use of them.

At this point we assume replica symmetry. We recall that while in
Ref. \cite{CFP} replica non-symmetric correlation functions were allowed, 
it was found that in the metastable state correlations are replica symmetric.
We consider only the metastable state and, thus, we restrict
ourselves to replica symmetric correlation functions. 
We comment further on this issue at the end of this Letter.
Specifically, we assume that all 
replica diagonal correlations are equal, $\delta n_{00}=\delta n_{11}=...=n^2 h$ and
$c_{00}=c_{11}=...=c$,  
as are all replica off-diagonal ones,
$\delta n_{01}=\delta n_{12}=...=n^2 \tilde{h}$ and 
$c_{01}=c_{12}=...=\tilde{c}$. Using the replica
symmetry we easily get the standard OZ equation
for the replica diagonal correlations,
\begin{equation}
\label{rOZaeqb}
h(\mathbf{r}_1,\mathbf{r}_2) = c(\mathbf{r}_1,\mathbf{r}_2)
+ n \int d\mathbf{r}_3 c(\mathbf{r}_1,\mathbf{r}_3)
h(\mathbf{r}_3,\mathbf{r}_2).
\end{equation}
Physically, this means that we have equilibrium 
correlations within each replica.

To get an equation for $\tilde{h}$ 
we first re-write the replicated OZ equation for
$\alpha\neq\beta$ in the following form:
\begin{eqnarray}
\label{rOZaneqb}
&& \int d\mathbf{r}_3 (\delta(r_{13})-n c_{\alpha\alpha}(\mathbf{r}_1,\mathbf{r}_3))
\delta n_{\alpha\beta}(\mathbf{r}_3,\mathbf{r}_2) 
\nonumber \\ &=& 
n \int d\mathbf{r}_3 c_{\alpha\beta}(\mathbf{r}_1,\mathbf{r}_3)
(n\delta(r_{32}) + \delta n_{\mathbf{\beta\beta}}(\mathbf{r}_3,\mathbf{r}_2))
\nonumber \\ && + 
n \sum_{\gamma\neq\alpha,\gamma\neq\beta} 
\int d\mathbf{r}_3 c_{\alpha\gamma}(\mathbf{r}_1,\mathbf{r}_3)
\delta n_{\gamma\beta}(\mathbf{r}_3,\mathbf{r}_2) 
\end{eqnarray}
In the $m\to 1$ limit from Eq. (\ref{rOZaneqb}) we obtain
\begin{eqnarray}
\label{rOZ01}
&& \int d\mathbf{r}_3 (\delta(r_{13})-n c(\mathbf{r}_1,\mathbf{r}_3))
\tilde{h}(\mathbf{r}_3,\mathbf{r}_2) = 
\tilde{c}(\mathbf{r}_1,\mathbf{r}_2)
\\ \nonumber &&
+ n \int d\mathbf{r}_3 \tilde{c}(\mathbf{r}_1,\mathbf{r}_3) 
h(\mathbf{r}_3,\mathbf{r}_2)
- n \int d\mathbf{r}_3 \tilde{c}(\mathbf{r}_1,\mathbf{r}_3)
\tilde{h}(\mathbf{r}_3,\mathbf{r}_2).
\end{eqnarray}
We note in passing that if we define a new replica off-diagonal correlation
function $\bar{h}$ through the following equation
\begin{eqnarray}
\label{newh}
\tilde{h}(\mathbf{r}_1,\mathbf{r}_2) &=& 
\int d\mathbf{r}_3 d\mathbf{r}_4 (\delta(r_{13})+n h(\mathbf{r}_1,\mathbf{r}_3))
\bar{h}(\mathbf{r}_3,\mathbf{r}_4)
\nonumber \\ && \times
(\delta(r_{42})+n h(\mathbf{r}_4,\mathbf{r}_2)),
\end{eqnarray}
we can derive from Eq. (\ref{rOZ01}) a somewhat simpler equation
for $\bar{h}$:
\begin{eqnarray}
\label{rOZ01a}
&& \bar{h}(\mathbf{r}_1,\mathbf{r}_2) = \tilde{c}(\mathbf{r}_1,\mathbf{r}_2)
\\ \nonumber &&
- n \int d\mathbf{r}_3 d\mathbf{r}_4 \tilde{c}(\mathbf{r}_1,\mathbf{r}_3)
(\delta(r_{34})+n h(\mathbf{r}_3,\mathbf{r}_4)) 
\bar{h}(\mathbf{r}_4,\mathbf{r}_2).
\end{eqnarray}
Eqs. (\ref{newh}-\ref{rOZ01a}) imply that 
replica off-diagonal correlation function $\tilde{h}$ can be written as a series 
in which the $k$th term is a convolution of the form 
$(-1)^{(k+1)} ((\delta+nh)\tilde{c})^k (\delta+nh)$. 

At this point we invoke the physically motivated relation \cite{PZ2} between
the Fourier transform of the replica off-diagonal correlation function, 
$\tilde{h}(k)$,
and the non-ergodicity parameter \cite{nep} of the mode-coupling theory, $f(k)$,
\begin{equation}
\label{htf}
n\tilde{h}(k) = S(k) f(k).
\end{equation}
Using Eq. (\ref{htf}) we can re-write the Fourier transform of Eq. (\ref{rOZ01}) as
\begin{eqnarray}
\label{rOZ01f}
\frac{f(k)}{1-f(k)} = n S(k) \tilde{c}(k).
\end{eqnarray}
We note the striking resemblance of Eq. (\ref{rOZ01f}) and the mode-coupling theory's
equation for the non-ergodicity parameter (see, \textit{e.g}, Eq. (2.17) of
Ref. \cite{BGS}) \cite{nepcomment}. Specifically, Eq. (\ref{rOZ01f}) 
coincides with the equation for the non-ergodicity parameter 
\textit{iff} the right-hand-side of this 
equation can be related to the long-time limit of the appropriate memory function.

We would like to emphasize that the derivation of Eq. (\ref{rOZ01f}) did not
involve any approximation. The only assumptions used were the existence of a metastable
state and the replica symmetry. We believe that Eq. (\ref{rOZ01f})
is the first result suggesting a connection between off-diagonal
direct correlation function in replica approach and the long-time limit of 
memory function introduced in dynamic approach. 

\section{Metastable state} 
The basic assumption of the replica approach is that if the 
density is high enough (or if the temperature is low enough) the system gets
stuck in a metastable state. In the original papers
this state was defined as a local minimum
of a free energy considered as a functional of the replica off-diagonal
correlations \cite{MP1,FP,CFP} or a function of the cage radius \cite{MP2,PZ1,PZ2}. 
Here we assume that the metastable state is
characterized by vanishing particle currents. We note that this condition is
qualitatively different than that used in earlier works in that it implicitly
depends on the system's dynamics. This is important since we assume 
metastability only with respect to local dynamics and we admit a possibility that
non-local \cite{GP} or cluster \cite{Krauth} 
moves can still equilibrate the system. For concreteness, 
we will here assume that the system is evolving with Brownian dynamics \cite{ND}. 
In this case we will see that the condition of vanishing particle currents
is equivalent to the assumption that the replica off-diagonal correlation
functions satisfy the first few equations of the 
Yvon-Born-Green (YBG) hierarchy \cite{HansenMcDonald}.

To motivate the assumption of vanishing particle currents we note that the 
time-independence of the joint density in replicas $\alpha$ and $\beta$, 
$\alpha\neq\beta$, results in the following identity
(to simplify notation we assume hereafter that 
the diffusion coefficient of an isolated particle has been made equal to 1 and that 
the inverse temperature $1/k_BT$ has been incorporated into the inter-particle
force $\mathbf{F}$):
\begin{eqnarray}
\label{timeindep}
0 &=& \partial_t n_{\alpha\beta}(\mathbf{r}_1,\mathbf{r}_2;t) 
\nonumber \\ &=&
-\partial_{\mathbf{r}_1} \mathbf{j}_{\alpha,\beta}(\mathbf{r}_1,\mathbf{r}_2;t)
-\partial_{\mathbf{r}_2} \mathbf{j}_{\beta,\alpha}(\mathbf{r}_1,\mathbf{r}_2;t)
\end{eqnarray}
where $\mathbf{j}_{\alpha,\beta}$ is the joint current density in replica $\alpha$
and particle density in replica $\beta$,
\begin{eqnarray}
\label{current1def}
\mathbf{j}_{\alpha,\beta}(\mathbf{r}_1,\mathbf{r}_3) &=& 
-\partial_{\mathbf{r}_1} n_{\alpha\beta}(\mathbf{r}_1,\mathbf{r}_3)
\\ \nonumber && 
+ \int d\mathbf{r}_2 \mathbf{F}(\mathbf{r}_{12})
n_{\alpha\alpha\beta}(\mathbf{r}_1,\mathbf{r}_2,\mathbf{r}_3)
\end{eqnarray}
and, correspondingly, $\mathbf{j}_{\beta,\alpha}$ is the joint current density 
in replica $\beta$ and particle density in replica $\alpha$. In Eq. (\ref{current1def})
$n_{\alpha\alpha\beta}$ is the joint two-particle density in replica $\alpha$ and
single-particle density in replica $\beta$.

The main fundamental assumption of the present approach is that not only 
$\partial_t n_{\alpha\beta}=0$ (which is required to have a stationary state) 
but moreover $\mathbf{j}_{\alpha,\beta}=0=\mathbf{j}_{\beta,\alpha}$. 

The vanishing of 
the joint current density in replica $\alpha$ and
particle density in replica $\beta$ leads to the following equation
relating $n_{\alpha\beta}$ and $n_{\alpha\alpha\beta}$:
\begin{eqnarray}
\label{current1}
-\partial_{\mathbf{r}_1} n_{\alpha\beta}(\mathbf{r}_1,\mathbf{r}_3)
+ \int d\mathbf{r}_2 \mathbf{F}(\mathbf{r}_{12})
n_{\alpha\alpha\beta}(\mathbf{r}_1,\mathbf{r}_2,\mathbf{r}_3) = 0.
\end{eqnarray}

In a very similar way, starting from the vanishing of 
the joint current density in replica $\beta$ and two-particle 
density in replica $\alpha$ we obtain an equation relating 
$n_{\alpha\alpha\beta}$ and the joint two-particle density in replicas $\alpha$
and $\beta$, $n_{\alpha\alpha\beta\beta}$:
\begin{eqnarray}
\label{current2}
%&& 
\mathbf{j}_{\beta,\alpha\alpha}(\mathbf{r}_3,\mathbf{r}_1,\mathbf{r}_2) &=& 
-\partial_{\mathbf{r}_3} n_{\alpha\alpha\beta}(\mathbf{r}_1,\mathbf{r}_2,\mathbf{r}_3)
\\ \nonumber &+&
\int d\mathbf{r}_4 \mathbf{F}(\mathbf{r}_{34})
n_{\alpha\alpha\beta\beta}(\mathbf{r}_1,\mathbf{r}_2,\mathbf{r}_3,\mathbf{r}_4) = 0.
\end{eqnarray}

We should emphasize here that equations (\ref{current1}-\ref{current2}) are valid 
only for $\alpha\neq\beta$ and, moreover, in the limit of vanishing pinning potential.
Furthermore, we should note that in this limit  
Eqs. (\ref{current1}-\ref{current2}) are identical to the YBG equations
for $n_{\alpha\beta}$ and $n_{\alpha\alpha\beta}$, $\alpha\neq\beta$.

Combining equations (\ref{current1}-\ref{current2}) we get
\begin{eqnarray}
\label{currentcomb}
&& \partial_{\mathbf{r}_1} \partial_{\mathbf{r}_3}
n^2 \tilde{h}(\mathbf{r}_1,\mathbf{r}_3) \equiv 
\partial_{\mathbf{r}_1} \partial_{\mathbf{r}_3}
n_{\alpha\beta}(\mathbf{r}_1,\mathbf{r}_3) 
\\ \nonumber &=& 
\int d\mathbf{r}_2 \mathbf{F}(\mathbf{r}_{12})
\int d\mathbf{r}_4 \mathbf{F}(\mathbf{r}_{34})
n_{\alpha\alpha\beta\beta}(\mathbf{r}_1,\mathbf{r}_2,\mathbf{r}_3,\mathbf{r}_4). 
\end{eqnarray}
The physical meaning of Eq. (\ref{currentcomb}) is that correlations
between particle densities in 
different replicas have to be accompanied by correlations of forces 
in these replicas.

It would be tempting to factorize the four-point
function at the right-hand-side of Eq. (\ref{currentcomb}). However, 
due to non-vanishing three-particle correlations, 
two-particle densities in replicas $\alpha$ and $\beta$
can be correlated \textit{via} single-particle density 
correlations. To take care of this, we
use the following decomposition of $n_{\alpha\alpha\beta\beta}$ which is
inspired by similar decompositions of equilibrium densities \cite{CA}:
\begin{floatequation}
\mbox{see Eq. (\ref{decomp})}
\end{floatequation}
\begin{widetext}
\begin{eqnarray}
\label{decomp}
&&
n_{\alpha\alpha\beta\beta}(\mathbf{r}_1,\mathbf{r}_2,\mathbf{r}_3,\mathbf{r}_4) = 
n_{\alpha\alpha}(\mathbf{r}_1,\mathbf{r}_2)n_{\beta\beta}(\mathbf{r}_3,\mathbf{r}_4) 
\nonumber \\ && + 
\sum_{\gamma\delta}\int d\mathbf{r}_5 d\mathbf{r}_6
\left<\delta n_{\alpha\alpha}(\mathbf{r}_1,\mathbf{r}_2)
\delta n_{\gamma}(\mathbf{r}_5)\right>
\left<\delta n_{\gamma}(\mathbf{r}_5) \delta n_{\delta}(\mathbf{r}_6)\right>^{-1}
\left<\delta n_{\delta}(\mathbf{r}_6) 
\delta n_{\beta\beta}(\mathbf{r}_3,\mathbf{r}_4)\right>
+ 
n^{\mathrm{irr}}_{\alpha\alpha\beta\beta}
(\mathbf{r}_1,\mathbf{r}_2,\mathbf{r}_3,\mathbf{r}_4)
\end{eqnarray}
\end{widetext}
The first term at the right-hand-side of Eq. (\ref{decomp}) does not contribute
to the force correlations. The second term describes all correlations 
between two-particle densities in replicas $\alpha$ and $\beta$ mediated
by single-particle correlations. In this term, the brackets $\left< ... \right>$
denote averaging over the replicated state,  
$\left<\delta n_{\alpha\alpha} \delta n_{\gamma}\right>$ is 
the correlation function of a fluctuation of a microscopic two-particle density 
in replica $\alpha$,
\begin{eqnarray}
\delta n_{\alpha\alpha}(\mathbf{r}_1,\mathbf{r}_2)
&=& \sum_{i\neq j} [ \delta(\mathbf{r}_1-\mathbf{r}_{i\alpha})
\delta(\mathbf{r}_2-\mathbf{r}_{j\alpha}) 
\nonumber \\ && - 
\left<\delta(\mathbf{r}_1-\mathbf{r}_{i\alpha})
\delta(\mathbf{r}_2-\mathbf{r}_{j\alpha})\right> ]
\end{eqnarray}
and a fluctuation of a microscopic 
single-particle density in replica $\gamma$, 
\begin{equation}
\delta n_{\gamma}(\mathbf{r}_1)
= \sum_i [ \delta(\mathbf{r}_1-\mathbf{r}_{i\gamma}) - 
\left<\delta(\mathbf{r}_1-\mathbf{r}_{i\gamma})\right> ],
\end{equation}
and $\left<\delta n_{\gamma} \delta n_{\delta}\right>^{-1}$
is the inverse of the correlation of single-particle density fluctuations,
\begin{equation}
\left<\delta n_{\gamma}(\mathbf{r}_5) \delta n_{\delta}(\mathbf{r}_6)\right>^{-1}
= \frac{\delta_{\gamma\delta}\delta(\mathbf{r}_5-\mathbf{r}_6)}{n} 
- c_{\gamma\delta}(\mathbf{r}_5,\mathbf{r}_6).
\end{equation}
Finally, the third term at the right-hand-side of Eq. (\ref{decomp}),
$n^{\mathrm{irr}}_{\alpha\alpha\beta\beta}$, contains correlations 
between two-particle densities in replicas $\alpha$ and $\beta$ that are 
mediated by at least two single-particle correlations. We should emphasize that 
Eq. (\ref{decomp}) is essentially a definition of 
$n^{\mathrm{irr}}_{\alpha\alpha\beta\beta}$.

Substituting decomposition (\ref{decomp}) into Eq. (\ref{currentcomb}) and then using
Eq. (\ref{current1}), the first equation of the standard 
Yvon-Born-Green hierarchy 
\cite{HansenMcDonald}, and the replicated OZ equations (\ref{rOZ}) 
we obtain the following relation
for the replica off-diagonal direct correlation function:
\begin{eqnarray}
\label{caneqb}
&& \partial_{\mathbf{r}_1} \partial_{\mathbf{r}_3}
n^2 \tilde{c}(\mathbf{r}_1,\mathbf{r}_3) \equiv
\partial_{\mathbf{r}_1} \partial_{\mathbf{r}_3}
n^2 c_{\alpha\beta}(\mathbf{r}_1,\mathbf{r}_3)  
\\ \nonumber &=& 
\int d\mathbf{r}_2 \mathbf{F}(\mathbf{r}_{12})
\int d\mathbf{r}_4 \mathbf{F}(\mathbf{r}_{34})
n_{\alpha\alpha\beta\beta}^{\mathrm{irr}}
(\mathbf{r}_1,\mathbf{r}_2,\mathbf{r}_3,\mathbf{r}_4) 
\end{eqnarray}
The only assumption used so far is the existence 
of a metastable state characterized by vanishing currents.

\section{Closure} 
Now we are in a position to introduce the main 
approximation: a factorization of the 
four-point correlation function $n_{\alpha\alpha\beta\beta}^{\mathrm{irr}}$.
For typical interactions (which are strongly repulsive
at short distances) a straightforward factorization 
would result in divergences. 
To arrive at an alternative factorization approximation
we note that Eqs. (\ref{newh}-\ref{rOZ01a}) suggest that replica
off-diagonal correlations contain, on each side, exact single 
replica correlations. Thus, to proceed we factor out the single-replica
analogs of $n_{\alpha\alpha\beta\beta}^{\mathrm{irr}}$ and
factorize only its remaining (middle) part:
\begin{floatequation}
\mbox{see Eq. (\ref{fact})}
\end{floatequation}
\begin{widetext}
\begin{eqnarray}
\label{fact}
n^{\mathrm{irr}}_{\alpha\alpha\beta\beta}
(\mathbf{r}_1,\mathbf{r}_2,\mathbf{r}_3,\mathbf{r}_4) 
&\approx &
\frac{1}{2^4} \int d\mathbf{r}_5 ... d\mathbf{r}_{12}
n^{\mathrm{irr}}_{22}(\mathbf{r}_1,\mathbf{r}_2,\mathbf{r}_5,\mathbf{r}_6)
\\ \nonumber && \times
\left[ n^{\mathrm{irr}}_{22}(\mathbf{r}_5,\mathbf{r}_6,\mathbf{r}_7,\mathbf{r}_8)^{-1}
n^{\mathrm{irr}}_{\alpha\alpha\beta\beta}
(\mathbf{r}_7,\mathbf{r}_8,\mathbf{r}_9,\mathbf{r}_{10})
n^{\mathrm{irr}}_{22}(\mathbf{r}_9,\mathbf{r}_{10},\mathbf{r}_{11},\mathbf{r}_{12})^{-1}
\right]_{\mathrm{fact}}
n^{\mathrm{irr}}_{22}(\mathbf{r}_{11},\mathbf{r}_{12},\mathbf{r}_3,\mathbf{r}_4)
\end{eqnarray}
\end{widetext}
Here $n^{\mathrm{irr}}_{22}$ is the single replica analog of 
$n^{\mathrm{irr}}_{\alpha\alpha\beta\beta}$ \cite{n22comment}, 
$(n^{\mathrm{irr}}_{22})^{-1}$ denotes
the inverse of $n^{\mathrm{irr}}_{22}$ and the brackets $[...]_{\mathrm{fact}}$
denote factorization of the expression in the brackets, \textit{e.g.},
\begin{equation}
[n^{\mathrm{irr}}_{\alpha\alpha\beta\beta}
(\mathbf{r}_1,\mathbf{r}_2,\mathbf{r}_3,\mathbf{r}_4)]
_{\mathrm{fact}}=n_{\alpha\beta}(\mathbf{r}_{13})
n_{\alpha\beta}(\mathbf{r}_{24})+(1\leftrightarrow 2).
\end{equation}

Explicitly, factorization approximation in Eq. (\ref{fact}) leads to the 
following expression for $n^{\mathrm{irr}}_{\alpha\alpha\beta\beta}$:
\begin{floatequation}
\mbox{see Eq. (\ref{factfin})}
\end{floatequation}
\begin{widetext}
\begin{eqnarray}
\label{factfin}
n^{\mathrm{irr}}_{\alpha\alpha\beta\beta}
(\mathbf{r}_1,\mathbf{r}_2,\mathbf{r}_3,\mathbf{r}_4) 
&\approx &
\frac{1}{2} \int d\mathbf{r}_5 ... d\mathbf{r}_{12}
n^{\mathrm{irr}}_{22}
(\mathbf{r}_1,\mathbf{r}_2;\mathbf{r}_5,\mathbf{r}_6)
\left( n^{-1} \delta(\mathbf{r}_{57})-c(r_{57})\right)
\left( n^{-1} \delta(\mathbf{r}_{68})-c(r_{68})\right)
\\ \nonumber && \times
n_{\alpha\beta}(\mathbf{r}_7,\mathbf{r}_9)
n_{\alpha\beta}(\mathbf{r}_8,\mathbf{r}_{10})
\left( n^{-1} \delta(\mathbf{r}_{9,11})-c(r_{9,11})\right)
\left( n^{-1} \delta(\mathbf{r}_{10,12})-c(r_{10,12})\right)
n^{\mathrm{irr}}_{22}
(\mathbf{r}_{11},\mathbf{r}_{12};\mathbf{r}_3,\mathbf{r}_4)
\end{eqnarray}
\end{widetext}

We now substitute (\ref{factfin}) into Eq. (\ref{caneqb}), take the Fourier transform, 
and follow the procedure used in the derivation of the so-called vertices of the 
mode-coupling theory \cite{SL} to obtain the following (approximate) identity
\begin{eqnarray}
\label{final}
&& \mathbf{k} \mathbf{k} \tilde{c}(k) = 
\frac{n^2}{2} 
\int \frac{d\mathbf{q}_1 d\mathbf{q}_2}{(2\pi)^3} 
\delta(\mathbf{k}-\mathbf{q}_1-\mathbf{q}_2)
\\ \nonumber && \times 
\left[ \mathbf{q}_1 c(q_1) + \mathbf{q}_2 c(q_2) \right]
\tilde{h}(q_1) \tilde{h}(q_2)
\left[ \mathbf{q}_1 c(q_1) + \mathbf{q}_2 c(q_2) \right].
\end{eqnarray}
We note that the bare forces in Eq. (\ref{caneqb}) got replaced by renormalized
forces (essentially, gradients of the replica-diagonal direct correlation functions).
Technically, this results from including $n^{\mathrm{irr}}_{22}$'s at both 
``ends'' of the approximate expression (\ref{factfin}).
 
Next, we take the longitudinal part of Eq. (\ref{final}), use the
identification (\ref{htf}) and in this way we derive an expression
for the replica off-diagonal direct correlation function $\tilde{c}$ in terms 
of the non-ergodicity parameter $f$,
\begin{eqnarray}
\label{finalfinal}
&& \tilde{c}(k) = 
\frac{1}{2k^2} 
\int \frac{d\mathbf{q}_1 d\mathbf{q}_2}{(2\pi)^3} 
\delta(\mathbf{k}-\mathbf{q}_1-\mathbf{q}_2)
\\ \nonumber && \times
f(q_1) f(q_2) S(q_1) S(q_2)
\left(\hat{\mathbf{k}}\cdot\left[ \mathbf{q}_1 c(q_1) + \mathbf{q}_2 c(q_2) 
\right]\right)^2
\end{eqnarray}
Finally, we substitute expression (\ref{finalfinal}) 
into Eq. (\ref{rOZ01f}). The resulting self-consistent equation for
the non-ergodicity parameter is identical to the equation derived using 
the standard mode-coupling theory. In particular, this equation implies the
following relation between $\tilde{c}$ and the long-time limit
of the irreducible memory function $M^{\mathrm{irr}}$,
\begin{equation}
\tilde{c}(k) = \frac{1}{n k^2} \lim_{t\to\infty} M^{\mathrm{irr}}(k;t).
\end{equation}

\section{Discussion}  
We have showed that the mode-coupling theory equation for the 
non-ergodicity parameter can be obtained by combining replica approach, 
condition of vanishing currents and the factorization approximation for the 
non-trivial part of the replica-off diagonal static four-point correlation function. 
One cannot but conclude that, at least as far as the calculation of the 
non-ergodicity parameter is concerned, 
the mode-coupling theory is equivalent to the (admittedly mean-field) 
replica approach. One could even say that in spirit it 
resembles the Kirkwood-Monroe theory of freezing \cite{KM}. 

Two features of the present development should be emphasized. First, 
the consistency of replica approach's equation for the off-diagonal 
correlations and the mode-coupling theory equation for the non-ergodicity parameter
is the strongest connection one could expect between the replica approach and the
mode-coupling theory. The reason is that the replica approach is an inherently 
static one and the bulk of mode-coupling theory's predictions are concerned with
dynamics. Second, since both replica approach and mode-coupling theory share 
mean-field character, it is natural to expect that both of them will involve
some factorization approximations. The advantage of the replica derivation 
presented here is that it relies upon a factorization approximation for static
correlation functions, albeit for a replicated system.

There are several aspects of our approach
that could be investigated further. First, a more accurate 
factorization approximation should be attempted. It can be argued that 
Eq. (\ref{factfin})
overestimates $n^{\mathrm{irr}}_{\alpha\alpha\beta\beta}$ and this leads to 
an underestimation of the mode-coupling transition density (or overestimation
of the mode-coupling transition temperature). Second, the limit of large
spatial dimensions should be investigated. In this limit static correlations
simplify and it is possible that factorization approximations could be avoided.
Third, replica symmetry 
should be inspected. In particular, one should investigate breaking symmetry
between the 0th replica and the remaining $r$ replicas. A  
preliminary study suggests that, at least for the hard-sphere interaction and with
Percus-Yevick closure for the replica diagonal correlations, this symmetry is not
broken.  Fourth, dependence of the final equation for the ergodicity breaking
parameter on the dynamics should be investigated. In particular, one could try
to derive the equation for the non-ergodicity parameter corresponding to
Monte Carlo dynamics and to check whether the standard equation
for the non-ergodicity parameter changes.

We thank G. Biroli and F. Zamponi for comments on the manuscript and
gratefully acknowledge the support of NSF Grant CHE 0909676.


\begin{thebibliography}{99}

\bibitem{BGS}
U. Bengtzelius, W. G\"otze, and A. Sj\"olander, J. Phys. C \textbf{17} 5915 (1984).

\bibitem{Goetze}
For a recent review see W. G\"otze,
\textit{Complex Dynamics of Glass-Forming Liquids: A Mode-Coupling Theory}
(Oxford University Press, Oxford, 2008).


\bibitem{MP1}
M. M\'ezard and G. Parisi, J. Phys. A: Math. Gen. \textbf{29} 6515 (1996).

\bibitem{FP} S. Franz and G. Parisi, Phys. Rev. Lett. \textbf{79}, 2486 (1997).

\bibitem{CFP}
M. Cardenas, S. Franz and G. Parisi, J. Chem. Phys. \textbf{110}, 1726 (1999).

\bibitem{MP2}
M. M\'ezard and G. Parisi, Phys. Rev. Lett. \textbf{82} (1999) 747; J. Chem. Phys.
\textbf{111} 1076 (1999).

\bibitem{critical}
M. Fixman, J. Chem. Phys. \textbf{36} 310 (1962); 
K. Kawasaki, Phys. Rev. \textbf{150} 291 (1996); 
L.P. Kadanoff and J. Swift, Phys. Rev. \textbf{166} 89 (1968). 

\bibitem{EHvL}
M.H. Ernst, E.H. Hauge and J. M. J. van Leeuwen, 
Phys. Rev. Lett. \textbf{25} 1254 (1970).

\bibitem{SL} For a derivation of mode-coupling theory for systems evolving with
overdapmed (Brownian) dynamics see 
G. Szamel and H. L\"{o}wen, Phys. Rev. A \textbf{44} 8215 (1991).

\bibitem{Hansen} J.N. Roux, J.L. Barrat, and J.P. Hansen,
J. Phys. Cond. Matt. \textbf{1}, 7171 (1989);
J.L. Barrat, J.N. Roux, and J.P. Hansen, Chem. Phys. \textbf{149}, 197 (1990).

\bibitem{nep} The non-ergodicity parameter $f(k)$ is defined in terms 
of the long-time limit of the intermediate scattering function $F(k;t)$, 
$f(k) = \lim_{t\to\infty} F(k;t)/S(k)$, where $S(k)$ denotes the static structure
factor, $S(k) = 1 + n h(k)$.

\bibitem{SK} F. Sciortino and W. Kob, Phys. Rev. Lett. \textbf{86}, 648 (2001).

\bibitem{PZ1} G. Parisi and F. Zamponi, J. Chem. Phys. \textbf{123}, 144501 (2005).

\bibitem{PZ2} G. Parisi and F. Zamponi, Rev. Mod. Phys. \textbf{82}, 789 (2010).

\bibitem{KTW} See T.R. Kirkpatrick and D. Thirumalai, Transp. Theor.
Stat. Phys. \textbf{24}, 927 (1995) and references therein. 

\bibitem{SS} B. Schmid and R. Schilling, Phys. Rev. E \textbf{81}, 041502 (2010).

\bibitem{IM} A. Ikeda and K. Miyazaki, Phys. Rev. Lett. \textbf{104}, 255704 (2010).

\bibitem{KW} The equation for the tagged particle localization length 
derived from a simplified mode-coupling theory
has been showed to coincide with the equation derived from the so-called
naive density-functional approach [T.R. Kirkpatrick and P.G. Wolynes,
Phys. Rev. A \textbf{35}, 3072 (1987)]. 

\bibitem{nepcomment} The resemblance of Eq. (\ref{rOZ01f}) and the 
mode-coupling theory's equation for the non-ergodicity parameter can 
be used to further justify the identification (\ref{htf}) of the replica off-diagonal
correlations and the non-ergodicity parameter.

\bibitem{GP} T. Grigera and G. Parisi, Pys. Rev. E \textbf{63}, 045102(R) (2001).

\bibitem{Krauth} L. Santen and W. Krauth, Nature \textbf{405}, 550 (2000).

\bibitem{ND} We anticipate that our procedure can be adapted to Newtonian 
dynamics with only minor additional assumptions.

\bibitem{HansenMcDonald}J.P. Hansen and I.R. McDonald, \textit{Theory of Simple Liquids}
(Elsevier, Amsterdam, 2006).

\bibitem{CA} J. B\l awzdziewicz, B. Cichocki and R. Ho\l yst, Physica A \textbf{157},
857 (1989); H.C. Andersen, J. Phys. Chem. B \textbf{106}, 8326 (2002).

\bibitem{n22comment} Strictly speaking, $n_{22}^{\mathrm{irr}}$ also includes
additional terms involving $\delta$ functions. Thus, the structure of 
Eq. (\ref{fact}) is analogous to that of Eq. (\ref{newh}).

\bibitem{KM} J.G. Kirkwood and E. Monroe, J. Chem. Phys. \textbf{9}, 514 (1941).


\end{thebibliography}
\end{document}